 \documentclass[final,5p,times,sort&compress,twocolumn]{elsarticle}

\usepackage{amssymb}
\usepackage[utf8]{inputenc}
\usepackage[T1]{fontenc}
\usepackage[hidelinks]{hyperref}
\DeclareUnicodeCharacter{2212}{-}
\journal{Physics Letters B (ADP-23-22/T1231)}

\begin{document}

\begin{frontmatter}
%\preprint{ADP-23-22/T1231}


\title{Excluded Volume Effects on Cold Neutron Star Phenomenology}


\author[first]{Jesper Leong}
\ead{jesper.leong@adelaide.edu.au}
\author[first]{Anthony W.~Thomas}
\author[second]{Pierre~A.~M.~Guichon}
\affiliation[first]{organization={CSSM and ARC Centre of Excellence for Dark Matter Particle Physics, Department of Physics, University of Adelaide},%Department and Organization
            postcode={5005}, 
            state={SA},
            country={Australia}}
\affiliation[second]{organization={Irfu, CEA, Universit\'e Paris-Saclay, F91191 Gif sur Yvette France}}            

\begin{abstract}
Observable properties of neutron stars are studied within a hadronic equation of state derived from the quark level. The effect of short-range repulsion is incorporated within the excluded volume framework. It is found that one can sustain neutron stars with masses as large as 2.2$M_\odot$ even including hyperons in $\beta$ equilibrium, while producing radii and tidal deformabilities consistent with current constraints. 
\end{abstract}




\end{frontmatter}

%\tableofcontents


\section{Introduction}
The past decade has seen a number of impressive advances in our knowledge of the properties of neutron stars (NS)~\cite{Lattimer:2006xb, Ozel:2016oaf, Motta:2022nlj, Steiner:2014pda,Martynov:2019gvu}, which have traditionally involved a degree of model 
dependence. We now know that there 
are stars with masses of order 2$M_\odot$~\cite{Antoniadis:2013pzd, NANOGrav:2017wvv, Riley:2021pdl} and possibly significantly 
higher~\cite{Kumar:2023mlp, Romani:2022jhd}. 
With the discovery of gravitational waves associated with the merger of two NS~\cite{LIGOScientific:2018cki, LIGOScientific:2017vwq, LIGOScientific:2020aai}, we not only have new constraints on their radii but information on a totally new property, the tidal deformability~\cite{Al-Mamun:2020vzu, Chatziioannou:2020pqz}. Satellite observations by the NICER collaboration~\cite{Riley:2021pdl} have provided new information on the radii of these fascinating objects~\cite{Bauswein:2017vtn,Miller:2021qha}. There has been a great deal of theoretical work related to these advances in order to test whether or not particular equations of state (EoS) are consistent with the new data~\cite{Raithel:2018ncd, Hotokezaka:2011dh, Chatziioannou:2020pqz, Chatziioannou:2015uea}. 

A special challenge for theorists has been the expected appearance of hyperons in heavy NS \cite{Burgio:2021vgk, Chatterjee:2015pua,Vidana:2000ew,Glendenning:1984jr}. In particular, for matter in $\beta$ equilibrium, as the central density of a NS increases one expects that the chemical potential of $\Lambda$ (and $\Xi^-$) hyperons must eventually equal the chemical potential of the neutrons (or neutron plus electron in the case of $\Xi^-$) \cite{Motta:2022nlj, Gal:2016boi}. As the momentum near the bottom of the hyperon Fermi sea is low, the appearance of hyperons slows the increase in pressure usually associated with an increase in energy density. In turn, the reduction in pressure makes it difficult for the star to sustain a large mass \cite{Burgio:2021vgk, Stone2007,Nishizaki:2002ih,Vidana:2000ew}. The observation of NS with masses in the region of 2$M_\odot$ was therefore a surprise, and often referred to as the ``hyperon crisis''. Although initially only the quark-meson coupling (QMC) model predicted NS with masses in this region containing 
hyperons~\cite{Stone2007}, there are now a number of models which do so, usually through the introduction of repulsive three-body forces \cite{Gomes:2014aka, Burgio:2021vgk, Chatterjee:2015pua}. However, such models struggle to produce NS containing hyperons with masses much higher than 2$M_\odot$.

Attempts to produce EoS that predict NS with masses above 2$M_\odot$, while straightforward if only nucleons are included, become problematic once hyperons play a role. This has led to the suggestion that there may be additional repulsion between the baryons at high density \cite{Harvey:1980rva, HALQCD:2018gyl}, to which nucleon-nucleon scattering data might not be sensitive \cite{Liu:1993sc}. It would be expected to play a role as the density rises and the degree of overlap between baryons increases. Hence it has been referred to as an ``overlap correction''~\cite{overlap}. This additional repulsion may arise through new forces beyond the Standard Model~\cite{Berryman:2021jjt,Berryman:2022zic}, or simply as the result of new arrangements of the quarks and gluons in multi-quark 
configurations~\cite{Inoue:2010hs,Harvey:1980rva}, with recent calculations finding that such an effect might increase the maximum NS mass ($M_{max}$) by several tenths of a solar mass \cite{overlap, Berryman:2021jjt}.

An alternate solution to the hyperon crisis is to posit a phase transition to deconfined quark 
matter~\cite{Masuda:2012kf,Buballa:2014jta,Baym:2019iky}, which requires interpolating between the well known low 
density~\cite{Gandolfi:2009fj, Kruger:2013kua} and high density (perturbative QCD) regions~\cite{Annala:2019puf, Kojo:2014rca, Kojo:2015nzn, Kurkela:2014vha, Motta:2022nlj, Whittenbury:2015ziz, Masuda:2012kf, Prakash:2021wpz,Most:2018eaw,Bentz:2002um}. This would require that the Gibbs free energy per baryon be lower for deconfined quark matter beyond some critical density~\cite{Suhonen:1987jc}. However, the threshold density and the EoS for deconfined quark matter at this region is poorly understood. Thus there is considerable model dependence associated with studies of this phase transition.

Early studies of ideal hadron gas models, used to simulate relativistic heavy ion collisions, were found to violate the Gibbs condition unless a correction referred to as the excluded volume effect (EVE) was incorporated~\cite{Suhonen:1987jc, Cleymans:1986cq}. By imposing a small exclusion volume around each baryon, the EVE correction models the suppression of the short distance relative wave functions induced by short range repulsion and the Pauli exclusion principle. This approach has since been shown to be both thermodynamically consistent and applicable to a more general class of phenomenological models~\cite{Rischke:1991ke}. Here we incorporate the EVE correction within the framework of the QMC and compare the predictions obtained from the resulting EoS against NS constraints.

In the QMC model the internal dynamics of an in-medium baryon is altered by the coupling of the exchanged mesons to the quarks themselves, naturally giving rise to many-body nuclear 
forces~\cite{Guichon:1987jp, Guichon:1989tx, Guichon:1995ue, Guichon:2018uew}, which involve no new parameters. The non-relativistic energy density functional derived within this model has shown some success in predicting the behaviour of finite nuclei with as few as 5 parameters~\cite{Martinez:2018xep, Martinez:2020ctv, Stone:2019syx, Stone:2017oqt, Saito:2005rv, Guichon:2018uew}. Early work using the relativistic formulation of QMC did predict the existence of NS with masses as large as $2M_\odot$, even when hyperonic matter was included~\cite{Guichon:2018uew, Stone2007}. However, like many relativistic models, it had a compressibility of nuclear matter that was somewhat large \cite{Motta:2019tjc, Miyatsu:2011bc, Whittenbury:2012rn}. In studies of finite nuclei it was found necessary to include a self-interaction term cubic in the $\sigma$ field~\cite{Martinez:2018xep, Guichon:2018uew}, in order to lower the incompressibility and match the energies of the giant monopole resonances~\cite{RevModPhys.89.015007, Sharma:2008uy, Piekarewicz:2002jd}. The strength of this additional term is governed by a non-zero coefficient, $\lambda_3$, which was found to be $\lambda_3 \in 0.02$-$0.05$ fm$^{-1}$~\cite{Martinez:2018xep}. Generally speaking, a higher incompressibility is linked to a stiffer EoS, which in turn leads to a higher $M_{max}$~\cite{Glendenning:1997wn}. Thus, this additional $\sigma$ self-interaction increases attraction, leading to a softer EoS, with the obvious consequence of reducing the $M_{max}$ the theory may predict~\cite{overlap}. As a result, in order to reach masses above $2M_\odot$ within a hadronic model, one needs additional repulsion.

As in many relativistic treatments, the short distance repulsion in the QMC model is generated by the  
$\omega$ meson, with the strength of its coupling set by nuclear matter properties at saturation density \cite{Guichon:2006}. The baryons are assumed to retain their individuality, with any potential overlap between the baryons ignored. At high densities, this assumption is expected to breakdown, and it is unlikely that the $\omega$ exchange alone will be able to capture the physics manifesting at supra-saturation densities. On the other hand, the EVE has a reasonable physical basis and is a well established theory used most commonly in the study of phase transitions to deconfined quark matter or chemical freeze-out in relation to data from relativistic heavy ion collisions~\cite{Suhonen:1987jc, Cleymans:1986cq, Braun-Munzinger:1999hun, Yen:1997rv}. Other applications include the study of possible van der Waals forces and the Mott 
effect~\cite{Ropke:1983lbc, Ropke:1984avu}. The baryons in the EVE are associated with a finite volume, which no other baryon is permitted to enter. The excluded volume, $v_0$, is governed by a single parameter, $r$. 

Panda {\em et al.} had previously applied the EVE correction to an earlier $\sigma$-$\omega$ QMC model and examined the in-medium changes in the hadron properties on the EoS of dense matter \cite{Panda:2002iu}. The authors also noted that the EVE EoS becomes notably stiffer with increasing hard core radius. However, they did not make a comparison with NS observations. The EVE is less commonly employed in NS studies because it is known to violate causality at very high density. Here we apply the excluded volume correction to the latest version of QMC and test it against heavy NS observations, including the tidal deformability, avoiding regions of density where this is a problem. 

%%%%%%%%%%%%%%%%%%%%%%%%%%%%%%%%%

\section{EVE Methodology and Results}
Rischke {\em et al.} gives a full outline of the EVE~\cite{Rischke:1991ke}. There are several ways to apply the EVE correction \cite{Typel:2016srf}, but here we choose the simplest version of the theory whereby each baryon is ascribed the same spherical volume $v_0=4\pi r^3/3$, parameterised by the exclusion radius, $r$. The excluded volume correction is not applied to the meson exchange particles nor the leptons. The key equations are presented in Eqs.~\ref{eq:density}-\ref{eq:EVEchem}. 
\begin{equation}
    \tilde{n}_i=\frac{n_i}{1-v_0n_B},
    \label{eq:density}
\end{equation}
\begin{equation}
    \tilde{\epsilon}(n_1, n_2,... ,n_i)=(1-v_0n_B)\epsilon(\tilde{n}_1, \tilde{n}_2,.., \tilde{n}_i),
    \label{eq:energy}
\end{equation}
\begin{equation}
    \tilde{P}(n_1, n_2,... ,n_i)=P(\tilde{n}_1, \tilde{n}_2,.., \tilde{n}_i),
    \label{eq:pressure}
\end{equation}
\begin{equation}
    \tilde{\mu}_j(n_1,n_2,...,n_i)=\mu_j(\tilde{n}_1,\tilde{n}_2,...,\tilde{n}_i)+v_0 P(\tilde{n}_1, \tilde{n}_2,...,\tilde{n}_i).
    %\tilde{\mu}(n)=\mu(\tilde{n})+v_0 P(\tilde{n}).
    \label{eq:EVEchem}
\end{equation}
In Eq.~\ref{eq:density}, the EVE individual baryon densities ($\tilde{n}_i$) are given in terms of the true individual baryon densities ($n_i$) and true total number density ($n_B$). To solve the EVE corrections at true densities, one substitutes the arguments in the QMC expressions with these EVE densities (see Refs.~\cite{Guichon:2018uew,overlap} for a complete summary of calculation of the relativistic QMC EoS used here). This is most clearly seen in Eq.~\ref{eq:pressure} where the EVE pressure ($\tilde{P}$) at true densities is computed using the definition of the QMC pressure ($P$) with $\tilde{n}_i$ as input. The EVE energy density ($\tilde{\epsilon}$) is computed in a similar fashion with a multiplicative factor of $1-v_0n_B$, as shown in Eq.~\ref{eq:energy}. Finally Eq.~\ref{eq:EVEchem} shows the correction to the chemical potential necessary for thermodynamic consistency.

As in recent studies of finite nuclei, there are five model parameters, with $m_\sigma=500$ MeV and $\lambda_3=0.05$ fm$^{-1}$ chosen and fixed throughout. The 3 remaining couplings (of the $\sigma$, $\omega$ and $\rho$ mesons) are fitted to reproduce $n_0=0.15$ fm$^{-3}$, $E_B/A=-15$ MeV, and $S=30$ MeV. Note that the binding energy per nucleon is defined as $E_B/A=\epsilon(n_B/2,n_B/2)/n_B-M_N$. If one fixes the couplings at $r=0$ fm, the minimum of the EVE binding energy, defined as $\tilde{E}_B/A=(1-v_0n_B)\epsilon(\tilde{n}_B/2,\tilde{n}_B/2)/n_B-M_N=\epsilon(\tilde{n}_B/2,\tilde{n}_B/2)/\tilde{n}_B-M_N$, will always be $-15$ MeV even when $r>0$. However, the saturation density, defined as the minimum of the $E_B/A$, will be less than $n_0=0.15$ fm$^{-3}$. To ensure the correct nuclear matter parameters, the couplings are refitted for each value of $r$. 

In the QMC model the bag radius, $R_B=1.0$ fm, defines the confinement region and is representative of the size of the nucleon. The radius of the excluded volume, $r$, is physically distinct from $R_B$, because it represents the distance at which the relative wave function is strongly suppressed~\cite{Panda:2002iu}. Clearly we expect $r<R_B$.  In this analysis the radius of the excluded volume is chosen to be $r=0,\ 0.4,\ 0.45$ and $0.5$ fm.

The refitted couplings, along with the incompressibility ($K$) and the slope of the symmetry energy ($L$), are presented in Table~\ref{tb:EVEcouplings}. Unlike the QMC model with a phenomenological overlap term~\cite{overlap}, the EVE correction does change the incompressibility and slope, both increasing as $r$ increases. For $r=0$ fm, $K=225$ MeV and this rises to $K=270$ MeV at $r=0.5$ fm, which is still within the experimental uncertainty ($200<K<300$ MeV)~\cite{Stone:2014wza, Blaizot:1980tw, Dutra:2012mb}. The trend in the incompressibility as $r$ increases is similar to that reported in Ref.~\cite{Panda:2002iu}, and one expects a stiffer EoS with larger $K$. $L$, which is observed to be $40<L<80$ MeV~\cite{Tsang:2012se}, only increases slightly, with a change of around $4$ MeV, over the same range of $r$. 
\begin{table}
\caption{\label{tb:EVEcouplings} The free space meson-nucleon couplings ($G_\sigma$, $G_\omega$ and $G_\rho$ with units of fm$^2$) are fitted for each choice of $r$ to reproduce  $n_0=0.15$ fm$^{-3}$, $E_B/A=-15$ MeV and $S=30$ MeV. $K$ and $L$, in units of MeV, show a moderate and slight increase with increasing $r$. }

    \centering
    \begin{tabular*}{\linewidth}{@{\extracolsep{\fill}} c ccc cc}
    \hline \hline
            r&$G_\sigma$&$G_\omega$&$G_\rho$ & K & L \\ \hline
            0 & 9.132 & 4.525 & 2.897 & 225 & 58 \\
            0.4 & 8.815 & 4.337 & 2.693 & 247 & 60 \\
            0.45 & 8.680 & 4.257 & 2.607 & 257 & 61 \\
            0.5 & 8.511 & 4.156 & 2.499 & 270 & 62 \\
            \hline \hline
    \end{tabular*}

\end{table}

In hadronic models of heavy NS hyperons are all important since their initial appearance, around $n_B \approx 3$$n_0$, softens the EoS. In Table~\ref{tb:EVEchem} we have listed the binding energies of the hyperons for completely symmetric nuclear matter at saturation density. These were calculated by subtracting the rest masses from EVE chemical potentials (Eq.~\ref{eq:EVEchem}). There is evidence that the $\Lambda$ in the 1s-state of Pb is bound by around $26$ MeV~\cite{Hashimoto:2006aw}. In comparison the $\Xi^-$ in $^{14}$N is weakly bound~\cite{Nakazawa:2015joa,Shyam:2019laf,Guichon:2008zz}. Our results for the 
$\Lambda$ and $\Xi$ reflect strong and weak attraction, respectively. These are the only hyperons which are predicted to appear in a NS at zero temperature (of course, at finite temperature the situation is far more 
complex~\cite{Stone:2019blq}), if we use the QMC EoS for matter in $\beta$ equilibrium. 
\begin{table}
\caption{\label{tb:EVEchem} In completely symmetric matter, the binding energies (MeV) for the nucleons ($N$) and hyperons ($\Lambda$ and $\Xi^{0,-}$) at saturation density are computed by subtracting the rest masses from the EVE chemical potentials given in Eq. \ref{eq:EVEchem}.}
\begin{tabular*}{\linewidth}{@{\extracolsep{\fill}}c c c c}
    \hline \hline
    r & $N$ & $\Lambda$ & $\Xi^{-,0}$ \\ \hline
    0 & -15 & -32 & -15 \\
    0.4 & -15 & -33 & -15 \\
    0.45 & -15 & -33 & -16 \\
    0.5 & -15 & -34 & -16 \\ 
    \hline \hline
    \end{tabular*}

\end{table}
\begin{eqnarray}
    \delta &\{ (1-v_0n_B)\epsilon_B(\tilde{n}_n, \tilde{n}_p, \tilde{n}_\Lambda, \tilde{n}_{\Xi^0}, \tilde{n}_{\Xi^-}) +\epsilon_l(e) + \epsilon_l(\mu) \nonumber \\ &
    + \nu_1\sum n_iq_i + \nu_2(\sum n_f - n_B) \}=0
    \label{eq:beta}
\end{eqnarray}

The type of NS stars considered here are long-lived and cold. At zero temperature, $\beta$ equilibrium is evoked and shown above in Eq.~\ref{eq:beta}. The first three terms minimise the EVE energy density. The final two terms include Lagrange multipliers ($\nu_{1,2}$), which act to conserve the total electric charge and baryon number. These are written in terms of the true number densities since conservation principles involve those. 
\begin{figure*}
    \centering
    \includegraphics[scale=0.7]{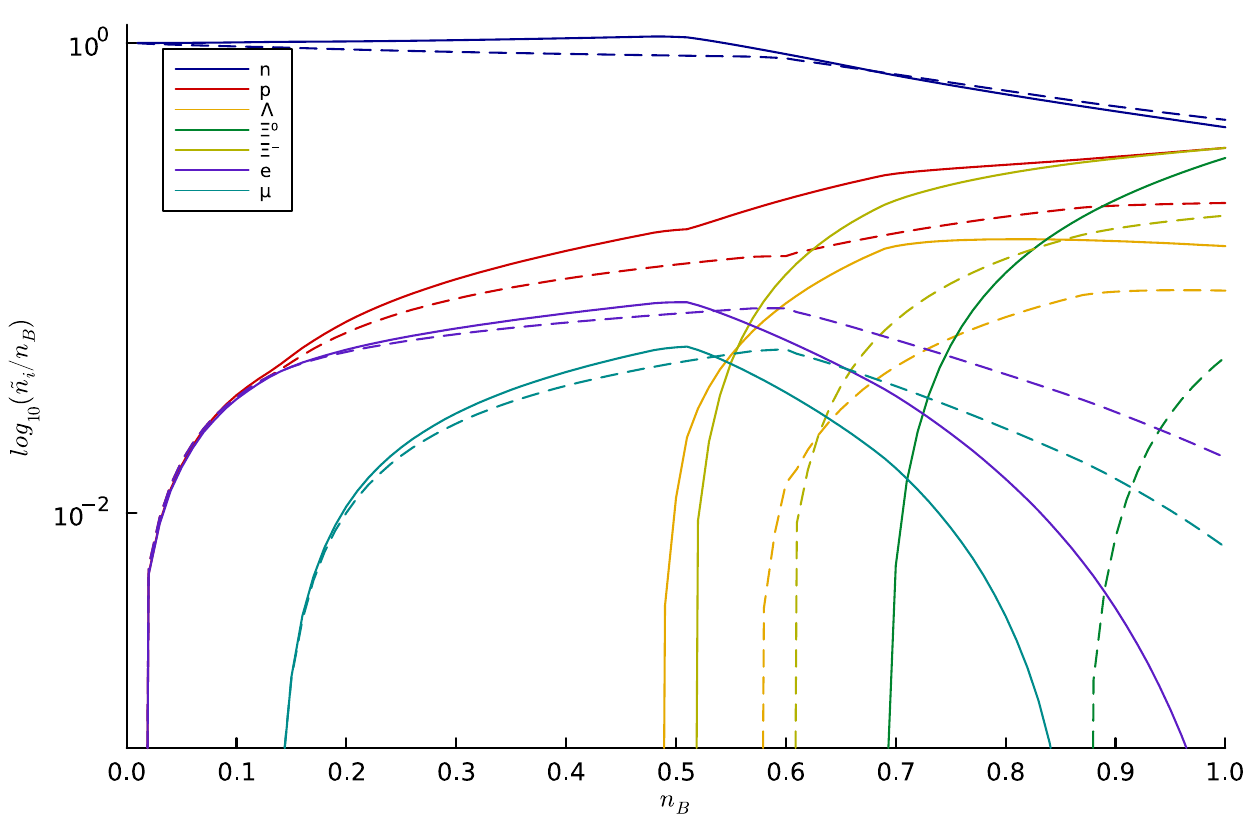}
    \caption{Species fraction relative to the true number density, $n_B$. The result for QMC with $r=0$ fm is shown with a dashed line, while $r=0.45$ fm is denoted by the solid lines. The hyperons appear slightly earlier with the EVE correction.}
    \label{fig:species}
\end{figure*}
%

%%%%%%%%%%%%%%%%%%%%%%%%%%%%
Figure~\ref{fig:species} shows the population of species fractions when $r=0$ fm (dashed) and $r=0.45$ fm (solid). As in previous studies, the $\Lambda$ appears before the $\Xi^-$, with the $\Xi^0$ appearing much later~\cite{Stone2007, overlap}. When the EVE is applied, the hyperons appear slightly earlier, but in the same order. No hyperons appear before $n_B = 3$$n_0$ and, as a result, stars with a core density lower than around $0.5$ fm$^{-3}$ are not populated by hyperons. Finally, to appropriately describe the NS, a low density crustal EoS provided by Hempel and Schaffner-Bielich~\cite{Hempel:2009mc, Hempel:2011mk} is matched to the QMC nuclear matter EoS at saturation density. 
\begin{figure*}
    \centering
    \includegraphics[scale=0.8]{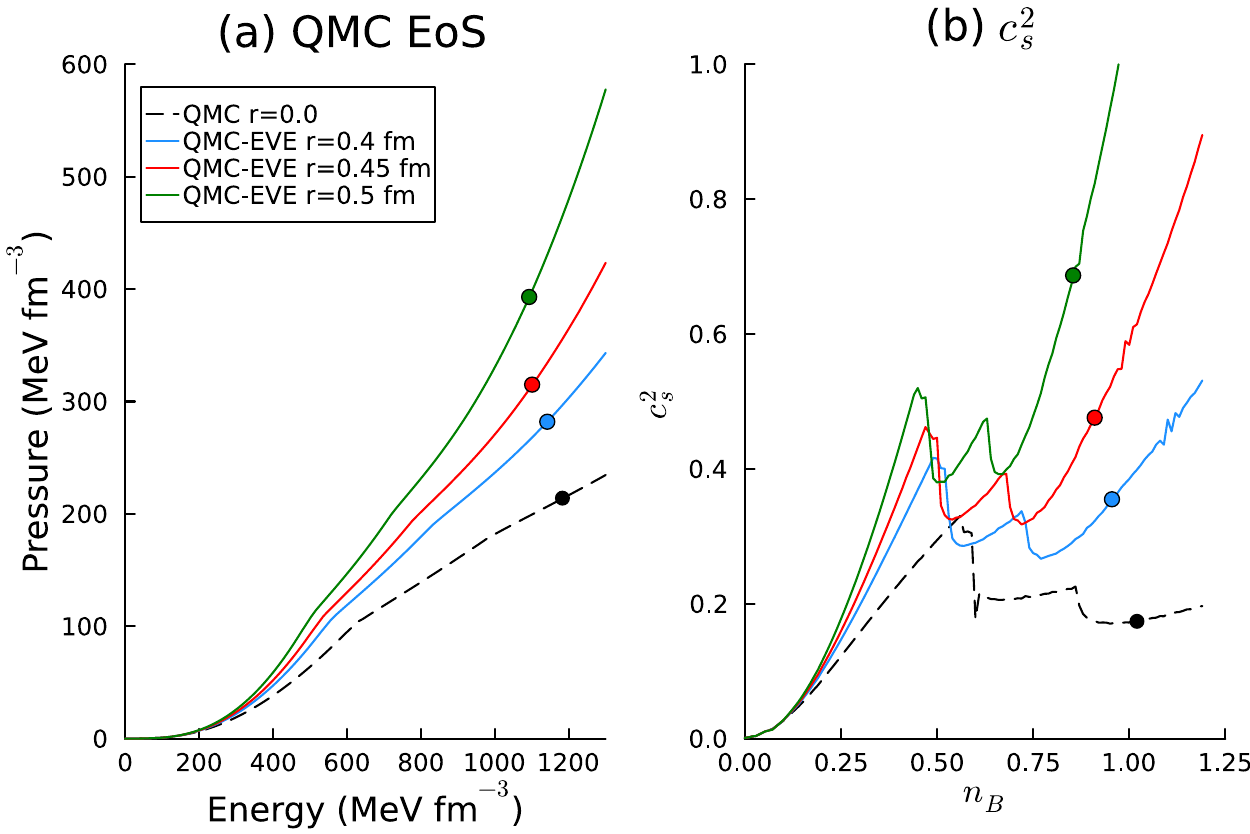}
    \caption{(a) QMC EoS with EVE correction for each choice of $r$ and (b) the corresponding $c_s^2$, speed of sound relative to light. Note that in (b) the true total number density is used. The markers indicate the central density, energy and pressure (shown in Table \ref{tb:EVECentral3}) at $M_{max}$. Sudden decreases in $c_s^2$ correspond to the appearance of hyperons.}
    \label{fig:EosSound}
\end{figure*}

The EVE is known to stiffen the EoS and one should be cautioned that the EoS will eventually become acausal for large $r$. The EoS is shown in Fig.~\ref{fig:EosSound}(a) and the corresponding $c_s^2$ is presented in Fig.~\ref{fig:EosSound}(b). $c_s^2=\frac{dP}{d\epsilon}$, with $c_s$ the speed of sound relative to the speed of light. Our results show a stiffening of the EoS, adding more pressure at higher densities as $r$ increases. Non-trivial losses in pressure correspond to the creation of hyperons, which carry little momentum upon initial appearance. This is readily seen by comparing the initial appearance of the hyperons in Fig.~\ref{fig:species} with sudden decreases in the $c_s^2$ in Fig.~\ref{fig:EosSound}(b). For $r=0.5$ fm, the EoS violates relativity just below $n_B=1.0$ fm$^{-3}$. Even for $r=0.45$ fm and $r=0.4$ fm, the EoS becomes acausal for large $n_B$. However, so long as the central density at $M_{max}$ is below the density at which special relativity is violated, as is the case here (see Table~\ref{tb:EVECentral3}), the EoS is valid. Markers displayed in Fig. \ref{fig:EosSound} indicated the central density, energy and pressure at $M_{max}$.

\begin{figure*}
    \centering
    \includegraphics[scale=0.7]{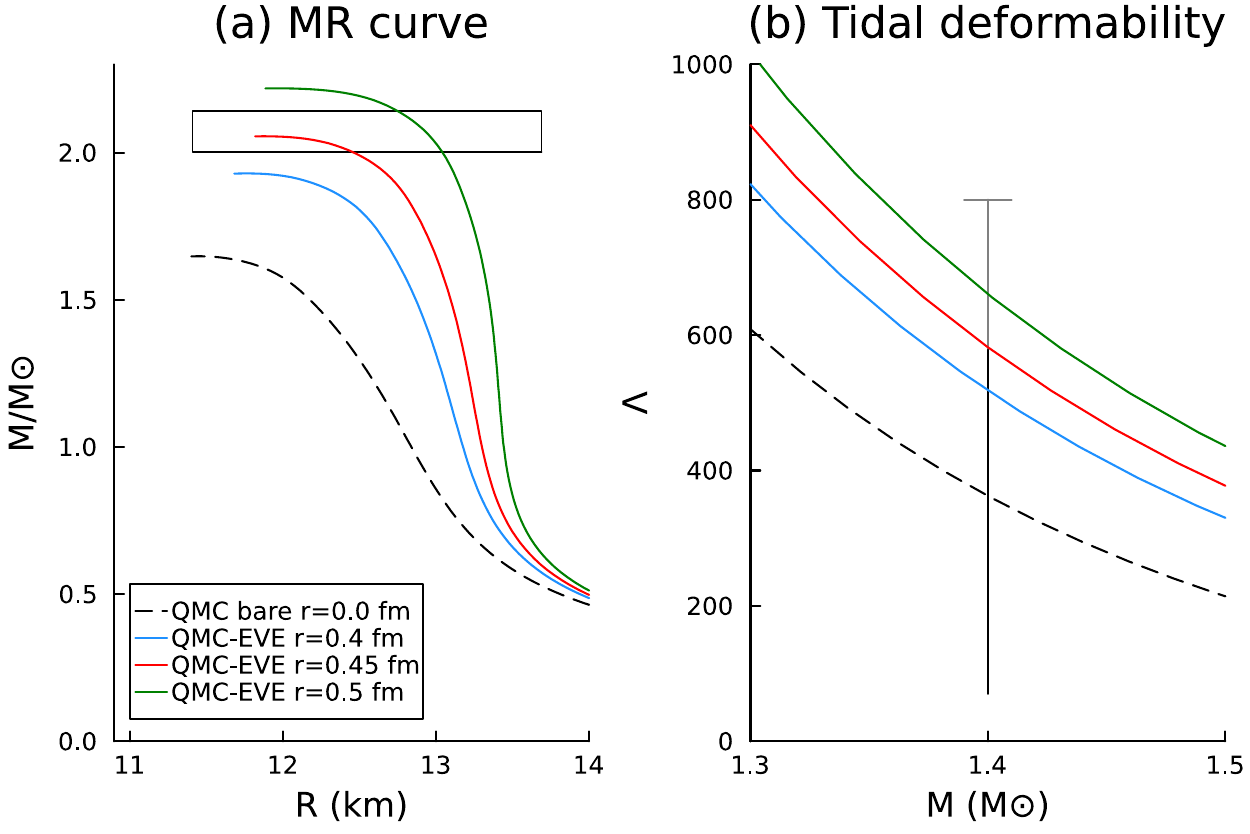}
    \caption{ (a) the mass-radius relationship of the NS derived from the QMC with EVE EoS. The observed heavy NS constraints are taken from Refs.~\cite{Riley:2021pdl, Fonseca:2021wxt}. (b) the tidal deformation for a $M=1.4M_\odot$ star. The constraint $\Lambda_{1.4}=190^{+390}_{-120}$ is indicated by the solid black line and is extracted from Ref.~\cite{LIGOScientific:2018cki}. Ref. \cite{LIGOScientific:2017vwq, Kim:2018aoi} suggest an upper limit of $\Lambda_{1.4}<800$, which is displayed here in grey.}
    \label{fig:MRTidal}
\end{figure*}

From the EoS the bulk properties of the NS are computed using the TOV equation. This assumes a spherical symmetric, non-rotating NS. The observed heavy NS star constraint is provided by  PSR J0740+6620 which one of the most massive and accurately measured NS known to date. It has a mass and radius of $M=2.072^{+0.067}_{-0.066}M{_\odot}$ and $R=12.39^{+1.30}_{-0.98}$ km \cite{Riley:2021pdl,Fonseca:2021wxt}. The one-sided tidal deformability~\cite{Hinderer:2009ca, Motta:2019tjc} provided by GW 170817 is $\Lambda_{1.4}=190^{+390}_{-120}$ \cite{LIGOScientific:2018cki}. Other studies suggest the tidal deformability may be as high as $\Lambda_{1.4}<800$ \cite{LIGOScientific:2017vwq, Kim:2018aoi}. The corresponding mass-radius  curves and tidal deformability are shown in Fig.~\ref{fig:MRTidal}.

\begin{table}
\caption{\label{tb:EVECentral3} The central number density ($n_c$, fm$^{-3}$), pressure ($\tilde{p}_c$, MeVfm$^{-3}$) and energy density ($\tilde{\epsilon}_c$, MeVfm$^{-3}$) for different mass stars ($M_\odot$) predicted by the QMC with EVE EoS. The last row for each case denotes the $M_{max}$ obtained. In addition, the radius (km) is recorded along with the total number of baryons (A, $10^{57}$).}
%\begin{tabular*}{\linewidth}{@{\extracolsep{\fill}} c ccc cc}
    \centering
    \begin{tabular*}{\linewidth}{@{\extracolsep{\fill}}cccccc}
    \hline \hline
    Mass & Radius & A & $n_c$ & $\tilde{p}_c$ & $\tilde{\epsilon}_c$  \\ \hline

        \multicolumn{6}{c}{$r=0.0$ fm} \\ 
    1.22 & 12.6 & 1.58 & 0.449 & 51 & 453 \\
    1.40 & 12.3 & 1.85 & 0.529 & 78 & 546 \\
    1.61 & 11.9 & 2.16 & 0.742 & 141 & 809 \\
    1.65 & 11.5 & 2.22 & 1.02 & 214 & 1183 \\ 

        \multicolumn{6}{c}{$r=0.4$ fm} \\
    1.21 & 13.1 & 1.57 & 0.378 & 41 & 378 \\
    1.41 & 12.9 & 1.86 & 0.430 & 59 & 436 \\
    1.61 & 12.8 & 2.15 & 0.489 & 85 & 506 \\
    1.82 & 12.5 & 2.47 & 0.608 & 135 & 656 \\
    1.93 & 11.7 & 2.66 & 0.955 & 282 & 1142 \\ 
    
    \multicolumn{6}{c}{$r=0.45$ fm} \\ 
    1.21 & 13.2 & 1.57 & 0.357 & 38 & 356 \\
    1.43 & 13.1 & 1.88 & 0.404 & 55 & 408 \\
    1.61 & 13.0 & 2.15 & 0.449 & 75 & 461 \\
    1.81 & 12.8 & 2.46 & 0.511 & 109 & 537 \\
    2.01 & 12.4 & 2.78 & 0.683 & 190 & 766 \\ 
    2.06 & 11.9 & 2.86 & 0.910 & 315 & 1101 \\ 
    
     \multicolumn{6}{c}{$r=0.5$ fm} \\ 
     1.20 & 13.4 & 1.56 & 0.333 & 34 & 330 \\
     1.40 & 13.4 & 1.84 & 0.369 & 47 & 370 \\
     1.60 & 13.3 & 2.14 & 0.408 & 65 & 415 \\
     1.80 & 13.2 & 2.44 & 0.451 & 90 & 467 \\
     2.00 & 13.0 & 2.76 & 0.522 & 130 & 557 \\
     2.20 & 12.4 & 3.10 & 0.755 & 272 & 892 \\%
     2.22 & 11.9 & 3.13 & 0.880 & 393 & 1093 \\
    \hline \hline
    \end{tabular*}

\end{table}
Given that the stiffness in the EoS corresponds to an increase in the $M_{max}$ of the predicted NS, we find that as the hard-core radius $r$ increases so to does the $M_{max}$ of the star. 
It is also clear that the star's radius also increases at all masses. We note that the low $M_{max}$ at $r=0$ fm is a consequence of choosing $\lambda_3=0.05$ fm$^{-1}$, which significantly lowers the incompressibility of nuclear matter.  

For incremental NS mass values, Table~\ref{tb:EVECentral3} shows the final total baryon number, radius, and central number density, pressure and energy density. The central density at $M_{max}$ when $r=0.5$ fm is $n_c=0.854$ fm$^{-3}$, which falls uncomfortably close, but below the density at which the EoS becomes acausal ($c_s^2=0.69$, see Fig. \ref{fig:EosSound}(b)). For $r=0.4$ fm and $r=0.45$ fm, the core density at $M_{max}$ corresponds to a speed of sound squared, $c_s^2=0.36$ and $c_s^2=0.48$, respectively. As also found in the overlap model~\cite{overlap}, the energy, pressure and number density increase quite fast as one moves from masses close to the maximum to the maximum, as seen from the last two rows of each permutation in Table~\ref{tb:EVECentral3}. For $r=0.45$ fm, the central density of a NS with a mass of $M=1.81M_\odot$ is high enough to allow the presence of $\Lambda$ hyperons. Below this mass, no hyperons are present, as can be seen from the species fraction graph in Fig.~\ref{fig:species}. For $M_{max}=2.06M_\odot$ the core contains the 
$\Lambda$, $\Xi^-$, and $\Xi^0$ hyperons. 

It has previously been suggested that the profile of $c_s^2$ with $n_B$ (see Fig.~\ref{fig:EosSound}(b)) offers clues to the composition of a NS. Annala {\em et al.} suggest that in heavy NS, $c_s^2<\frac{1}{3}$ is a signature of sizable deconfined quark matter core~\cite{Annala:2019puf}.
Motta {\em et al.}, working with the QMC model, challenged this interpretation, showing that this is not a unique feature of deconfined quark matter. In particular, purely hadronic models with hyperons present a similar $c_s^2$ profile~\cite{Motta:2020xsg}. Our results for $r \geq 0.45$ fm predict a $M_{max}$ greater than $2.0M_\odot$, even when hyperons are included. As we see in Fig.~\ref{fig:EosSound}(b), at the very highest masses (corresponding to the highest central densities) the EVE yields $c_s^2>\frac{1}{3}$, which does violate the conformal bound suggested by Annala {\em et al.}~\cite{Annala:2019puf} and constitutes a heavy neutron star core with only hadronic matter. This is not surprising since QMC is a hadronic model and contains no elements of deconfined quark matter but is nevertheless constrained by NS observations. We note that the conformal bound of $\frac{1}{3}$ is derived in the high temperature limit~\cite{Cherman:2009tw} and has not been shown to apply to the zero-temperature, high density regime. The attractiveness of the conformal limit is to provide a useful guide to interpolating well known low density and high density perturbative QCD EoS. But this may not be immediately necessary as Refs. \cite{Kojo:2015nzn} and \cite{Kurkela:2014vha} present interpolated EoS with quark matter cores strongly violating the conformal limit. Clearly much work stills needs to be done if the speed of sound, often used to test causality, is to be used to described the composition of the heaviest NS. There is a distinct possibility that the composition of NS may be experimentally clarified by future observations of  gravitational waves~\cite{Chatziioannou:2015uea, Chatziioannou:2020pqz, Bauswein:2018bma}.

Around the canonical mass value of $M \in 1.4$-$1.6M_\odot$, the radius is thought to be $R_{1.4}<13$ km~\cite{Ozel:2016oaf, Motta:2022nlj, Oertel:2016bki}. Some estimates find the experimental bounds to be even lower, in the range $10.1<R_{1.5}<11.1$ km~\cite{Oertel:2016bki}. Our results show that a $M=1.4M_\odot$ star has a radius between 12.3 and 13.4 km (see 
Table~\ref{tb:EVECentral3}). Given the large uncertainties and model dependency, much hope is directed towards tidal deformability and future binary NS merger events, since the tidal deformability is highly sensitive to the radius of the star (e.g., see Refs.~\cite{Hinderer:2009ca, Motta:2019tjc, Kalaitzis:2019dqc}). So far there has only been one confirmed twin binary neutron star merger event, GW 170817~\cite{LIGOScientific:2018cki}, but this number is expected to increase significantly in the near future. The computed tidal deformation for QMC with the EVE correction is presented in 
Fig.~\ref{fig:MRTidal}(b). We find that for $r=0.45$ fm the computed $\Lambda_{1.4}$ lies just at the boundary of the present experimental constraint provided by Ref.~\cite{LIGOScientific:2018cki}. For $r=0.5$ fm the tidal deformability lies outside this constraint.
Of course, measurements of the tidal deformation are still in their infancy, with another analysis suggesting that the GW 170817 data is consistent with $\Lambda_{1.4}<800$ as an upper-bound~\cite{Kim:2018aoi, LIGOScientific:2017vwq}. In that case the choice $r=0.5$ fm would not be ruled out by the tidal deformability constraint and the $M_{max}$ that the theory can sustain could be as high as 2.22 M$_\odot$.

%%%%%%%%%%%%%%%%%%%%%%%%%%%%%%%%%%

\section{Conclusion}
\label{sec:conclusion}
Relativistic EoS tend to produce values for the incompressibility of nuclear matter that are significantly higher than those produced by non-relativistic calculations based upon the same underlying physics. However, calculations within the QMC model using a non-relativistic energy density functional derived from the relativistic formulation require an additional term cubic in the scalar field in order to reproduce the observed giant monopole resonance energies. This additional term in the Lagrangian density has the effect of reducing the incompressibility of nuclear matter when introduced into the relativistic EoS calculations and that leads to a lower $M_{max}$.

Since we now have proof of the existence of NS with masses as large as 2.1 M$_\odot$, we have been led to search for the effects of physics that has so far been overlooked. The excluded volume formalism provides a simple and effective way to move beyond simple mean field theory by parametrizing the effect of short-range correlations. Each nucleon has a volume of radius $r$ associated with it from which other baryons are excluded. We have found that the choice $r=0.45$ fm is consistent with both the mass of the heaviest NS and the tidal deformability deduced from the gravitational wave observation GW170817. The highest mass found using the QMC EoS when hyperons are included is 2.2 M$_\odot$, which is obtained with $r=0.5$ fm; although that is in mild tension with some estimates of the upper bound on the tidal deformability $\Lambda_{1.4}$. As any larger choice of $r$ would lead to an acausal speed of sound at the densities reached in the heaviest stars, it appears very difficult to generate NS much heavier than 2.2 M$_\odot$, which are composed of baryons in $\beta$-equilibrium.
\section*{Declaration of competing interest}
The authors declare that they have no known competiting financial interests or personal relationships that could have appeared to influence the work reported in this paper.
\section*{Data availability}
Data will be made available upon request.
\section*{Acknowledgements}
We are pleased to acknowledge the contributions from Theo F. Motta in developing and assisting in the calculations for this paper. This work was supported by the University of Adelaide and the Australian Research Council through a grant to the ARC Centre of Excellence for Dark Matter Particle Physics (CE200100008).

\bibliographystyle{elsarticle-num} 
\bibliography{apssamp}



%% For example:

%% \bibitem[Aladro et al.(2015)]{Aladro15} Aladro, R., Martín, S., Riquelme, D., et al. 2015, \aas, 579, A101

%%\end{thebibliography}

\end{document}